\documentclass[printer]{aa}

\newcommand{\comb}[1]{{#1}\/}   
\newcommand{\out}[1]{}  

\newcommand{\nnh}{\hbox{$n_{\rm H}$}}
\newcommand{\phih}{\hbox{$\varphi_{\rm H}$}}
\newcommand{\map}{\hbox{{\sc mappings i}c}}
\newcommand{\up}{\hbox{\it U}}

\newcommand{\phn}{\hbox{~}}

\newcommand{\cmc}{\hbox{${\rm cm^{-3}}$}}
\newcommand{\kms}{\hbox{${\rm km\,s^{-1}}$}}

\newcommand{\llu}{\hbox{${\rm erg\, cm^{-2}\, s^{-1}}$}}

\newcommand{\nh}{\hbox{$n_{H}$}}
\newcommand{\Vs}{\hbox{$V_{s}$}}

\newcommand{\anu}{\hbox{$\alpha_{\nu}$}}

\newcommand{\aox}{\hbox{$\alpha_{OX}$}}

\newcommand{\zq}{\hbox{$z_q$}}
\newcommand{\kev}{\hbox{\sl keV}}

\newcommand{\II}{\hbox{\sc ii}}

\newcommand{\IV}{\hbox{\sc iv}}

\newcommand{\OI}{\hbox{O\,{\sc i}}}
\newcommand{\OII}{\hbox{O\,{\sc ii}}}
\newcommand{\OIII}{\hbox{O\,{\sc iii}}}
\newcommand{\FeIII}{\hbox{Fe\,{\sc iii}}}

\newcommand{\HeI}{\hbox{He\,{\sc i}}}
\newcommand{\HeII}{\hbox{He\,{\sc ii}}}
\newcommand{\NeVIII}{\hbox{Ne\,{\sc viii}}}
\newcommand{\MgX}{\hbox{Mg\,{\sc x}}}

\newcommand{\OIV}{\hbox{O\,{\sc iv}}}
\newcommand{\OV}{\hbox{O\,{\sc v}}}
\newcommand{\OVI}{\hbox{O\,{\sc vi}}}
\newcommand{\CaII}{\hbox{Ca\,{\sc ii}}}
\newcommand{\FeII}{\hbox{Fe\,{\sc ii}}}
\newcommand{\SiII}{\hbox{Si\,{\sc ii}}}
\newcommand{\SiIII}{\hbox{Si\,{\sc iii}}}
\newcommand{\SiIV}{\hbox{Si\,{\sc iv}}}
\newcommand{\SIII}{\hbox{S\,{\sc iii}}}
\newcommand{\SIV}{\hbox{S\,{\sc iv}}}
\newcommand{\NIV}{\hbox{N\,{\sc iv}}}
\newcommand{\NII}{\hbox{N\,{\sc ii}}}
\newcommand{\NIII}{\hbox{N\,{\sc iii}}}
\newcommand{\NV}{\hbox{N\,{\sc v}}}
\newcommand{\CIV}{\hbox{C\,{\sc iv}}}

\newcommand{\CIII}{\hbox{C\,{\sc iii}}}
\newcommand{\CII}{\hbox{C\,{\sc ii}}}
\newcommand{\w}{\hbox{$\lambda$}}
\newcommand{\OVIw}{\hbox{O\,{\sc vi}\,$\lambda 1035$}}

\newcommand{\SiIVw}{\hbox{Si\,{\sc iv}\,$\lambda 1400$}}
\newcommand{\CIVw}{\hbox{C\,{\sc iv}\,$\lambda 1549$}}
\newcommand{\CIIIw}{\hbox{C\,{\sc iii}\,$\lambda 977$}}

\newcommand{\ton}{\hbox{Ton\,34}}

\newcommand{\Hb}{\hbox{H$\beta$}}
\newcommand{\Lya}{\hbox{Ly$\alpha$}}
\newcommand{\Lyb}{\hbox{Ly$\beta$}}

\newcommand{\nhz}{\hbox{$n^0_{\rm H}$}}

\newcommand{\NN}{\hbox{$N_{20}^{\rm H}$}}

\newcommand{\sed}{\hbox{{\sc sed}}}
\newcommand{\seds}{\hbox{{\sc sed}{\rm s}}}
\newcommand{\Fnu}{\hbox{F$_\nu$}}

\newcommand{\Fla}{\hbox{F$_\lambda$}}

\usepackage{graphics}

\begin{document}



\title{The emission line spectrum of the UV deficient quasar Ton\,34: \\evidence of shock excitation?}


\author{Luc Binette\inst{1,2} and Yair Krongold\inst{2}}

\authorrunning{Binette et~al.}


\institute{D\'{e}partement de Physique, de G\'{e}nie Physique et d'Optique,
Universit\'{e} Laval, Qu\'{e}bec, QC, G1K\,7P4 \and Instituto de
Astronom\'\i a, UNAM,  Ap. 70-264, 04510 M\'exico, DF, M\'exico}
\date{Received: 12th July 2007/ Accepted: 20th September 2007 }

\titlerunning{The emission lines of \ton}

\authorrunning{Binette  \& Krongold}

\abstract{Emission lines in quasars are believed to originate from a
photoionized plasma. There are, however, some emission features
which appear to be collisionally excited, such as the \FeII\
multiplet bands. Shortward of \Lya, there also are a few permitted
lines of species from low to intermediate ionization.} {\ton\
($\zq=1.928$) exhibits the steepest far-UV continuum decline known
($\Fnu \propto \nu^{-5.3}$) shortward of 1050\,\AA. This object also
emits unusually strong low to intermediate excitation permitted
lines shortward of the Lyman limit.}{Using archive spectra of \ton\
from HST, IUE and Palomar, we measure the fluxes of all the lines
present in the spectra and compare their relative intensities with
those observed in composite quasar spectra.} {Our analysis  reveals
unusual strengths with respect to \Lya\ of the following low to
intermediate excitation permitted lines: \OII+\OIII\ (835\,\AA),
\NIII+\OIII\ (686--703\,\AA) and \NIII+\NIV\ (765\,\AA). We compare
the observed line spectrum with both photoionization and shock
models.}{Photoionization cannot reproduce the strengths of these
far-UV lines. Shocks with $\Vs \simeq 100$\,\kms\ turn out to be
extremely efficient emitters of these lines and are favored as
excitation mechanism. }


\maketitle



\section{Introduction} \label{sec:intro}

In this work, we analyze the emission lines of an unusual quasar,
\ton, which is alternatively named PG\,1017+280 or J1019+2745 with
redshift $\zq=1.928$. It is severely deficient in ionizing photons,
since its Spectral Energy Distribution (\sed) shows a remarkable
steepening of the continuum in the rest-frame far-UV, shortward of
1100\,\AA\ (Binette \& Krongold 2007, hereafter BK07; Binette
et\,al. 2007). If the far-UV is fitted by a powerlaw ($\Fnu \propto
\nu^{+\alpha}$), the index\footnote{Among the 77 quasars whose
far-UV indices could be measured by Telfer et\,al. 2002, there were
only 3 objects with a continuum steeper than $\nu^{-3}$.} is as
steep as $\nu^{-5.3}$. BK07 suggest that the extreme-UV flux might
undergo a recovery shortward of 450\,\AA.

While the near-UV emission-line spectrum appears to be `normal', the
far-UV spectrum shows low to intermediate ionization species with
unusual strengths. Using the UV \sed\ constructed by BK07 from
archive data, we will quantify this statement and present
photoionization and shock models for comparison. The aim is to
understand how the extreme deficiency of ionizing photons in \ton\
might be impacting the emission line spectrum.

The emission-line spectrum of quasar and Seyfert\,I galaxies is
generally believed to originate from gas photoionized by a nuclear
UV source. State of the art photoionization models of the Broad
Emission Line Region (BELR) such as those developed by Baldwin
et\,al. (1995) and dubbed `locally optimally emitting clouds' (LOC)
models can successfully reproduce most of the emission lines
observed in quasars. A grid of such models can be found in Korista
et\,al. (1997, hereafter KO97) and more recently in Casebeer et\,al.
(2006 and references therein). There are, however,  a few exceptions
to the success of pure photoionization. In particular,
photoionization models require micro-turbulences in order to
reproduce the shape and intensity of the \FeII\ UV-band (Baldwin
et\,al. 2004). A possible alternative is that the region producing
\FeII\ is collisionally ionized, as proposed by Grandi (1981, 1982),
Joly (1987), V\'eron-Cetty et\,al. (2004, 2006) and Joly et\,al.
(2007). In this work, we present evidence that photoionization might
not be sustainable in the case of some of the far-UV permitted lines
reported in this paper.

\section{The UV emission line spectrum of  \ton}\label{sec:lin}

Below we summarize the procedure used by BK07 to derive the UV \sed\
of \ton.

\subsection{Description of the archival data}\label{sec:spc}

The current work is based on four archival or bibliographical
sources. The 760--1120\,\AA\ spectral segment is provided by the
dataset Y2IE0A0AT from the HST-FOS archives (grating G270H). To
cover the extreme UV region, we borrowed from the IUE archives. The
long wavelength segment (LWP) is from Tripp, Bechtold \& Green
(1994) and corresponds to the dataset LW0P5708.  Fluxes longward of
3000\,\AA\ (observer-frame) were severely affected from reflected
sunlight or moonlight (Lanzetta, Turnshek \& Sandoval 1993) and have
been discarded. The shorter wavelength IUE segment (SWP) was
extracted directly from the archives and corresponds to the dataset
SWP28188.  To cover the \sed\ behavior longward of the HST segment,
we adopted the published optical spectra of Sargent, Boksenberg \&
Steidel (1988), which were taken at the Palomar 5.08\,m Hale
Telescope.  Both optical spectra lacked absolute flux calibration,
although the authors observed standard stars, which allowed them to
provide a relative calibration.

\subsection{Matching the different \sed\ segments}  \label{sec:sed}

We statistically corrected the UV spectral segments for the
cumulated absorption caused by unresolved \Lya\ forest lines,  which
are responsible for the so-called far-UV ``Lyman valley'' (M{\o}ller \&
Jakobsen 1990). For that purpose, we adopted the
\emph{mean}\footnote{This correction is statistical in nature, as it
relies on the average behavior with redshift of the spatial density
of intervening absorbers. It cannot be used to correct small
portions of the continuum, which may be coincident with a ``clear
patch'' or an over-density in the \Lya\ forest. These
inhomogeneities may generate spurious narrow features that  should
not be attributed to genuine emission lines.} transmission function
for $\zq=2$ published by Zheng et\,al. (1997). We also applied a
Galactic reddening correction assuming the Cardelli, Clayton \&
Mathis (1989) extinction curve corresponding to ${\rm R}_{\rm
V}=3.1$ and ${\rm E}_{B-V} = 0.13$. The latter value corresponds to
the mean extinction inferred  from the 100$\,\mu$ maps of Schlegel
et\,al. (1998) near \ton.  The blue and red arm segments have been
scaled to overlap smoothly with the HST-FOS segment. Both the LWP
and SWP segments were multiplied by a factor 0.75. This scaling was
necessary so that the LWP segment superimposes the HST-FOS spectrum
as closely as possible. Continuum variability is a possible
explanation for this continuum difference, since the IUE and HST
observations were made in different years. Finally, all the spectral
segments were shifted to rest-frame wavelengths, and \Fla\ was
multiplied by $1+\zq$. The IUE spectra have been re-binned by
grouping $n$ pixels together [SWP with $n=5$ and LWP with $n=3$) in
order to improve the limited S/N. The LWP and HST-FOS spectra
overlap significantly in spectral coverage. \comb{Both datasets
taken nine year apart confirm the unusual steepness of the UV break
in \ton.}

\begin{figure}
\resizebox{\hsize}{!}{\includegraphics{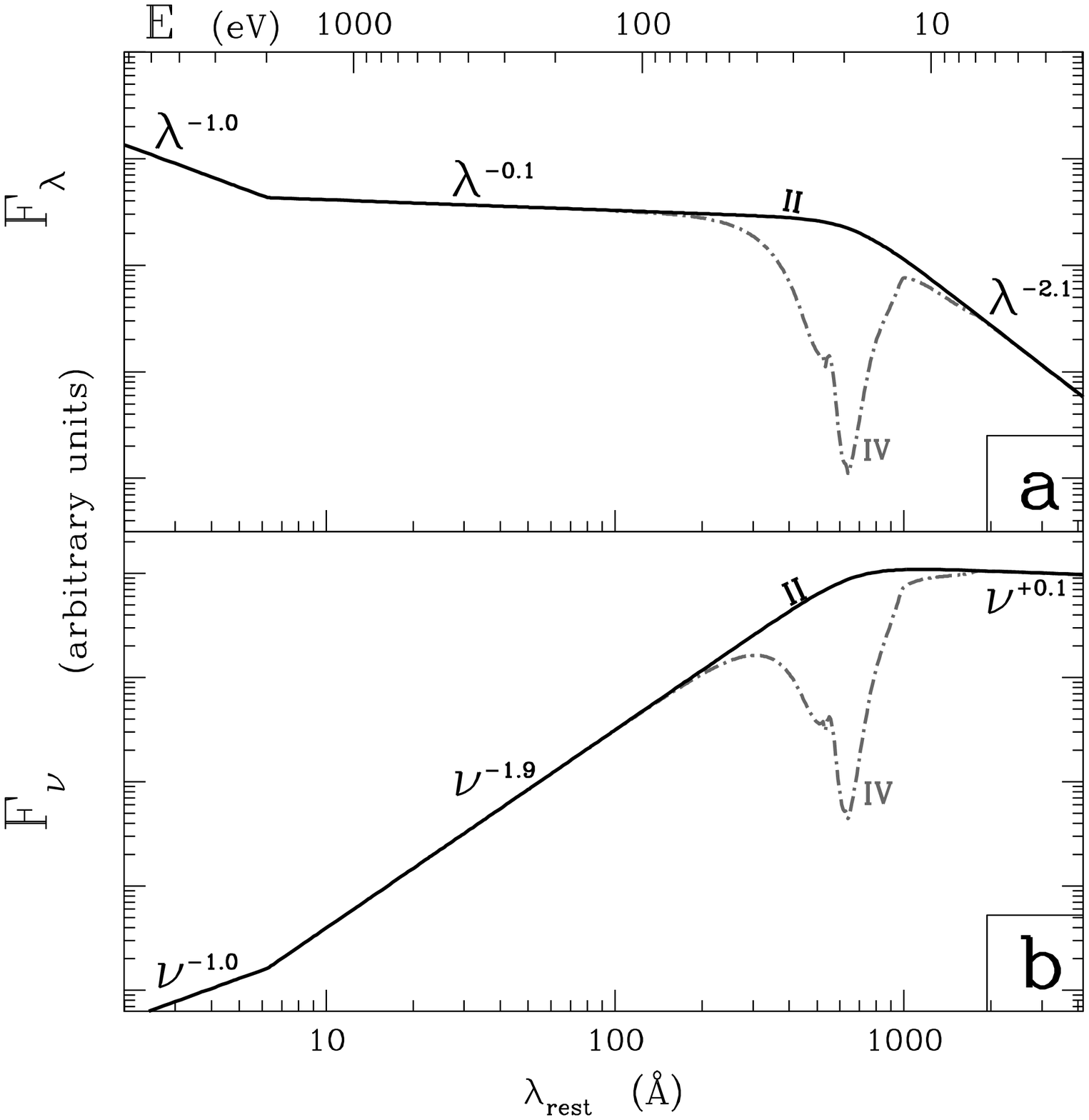}}
\caption{Log-log plot of the input spectral energy distributions
used in our photoionization calculations discussed in
Sect.\,\ref{sec:pho}.  These ionizing \seds\ are labeled \II\ and
\IV\ in either \Fla\ (panel {\it a}) or \Fnu\ (panel {\it b}) and
are given as a function of wavelength (bottom axis) or photon energy
(top axis). The distribution labeled \II\ (solid line) is the
assumed intrinsic \sed\ while that labeled \IV\ is the {\sl
transmitted} flux (dash-dotted line) assuming nanodiamond dust
extinction (see BK07).  Model\,\IV\  is a fit  of the UV continuum
of \ton\ between 400 and 1550\,\AA. \label{fig:sed}}
\end{figure}

\subsection{Model of the ionizing \sed\ of \ton}\label{sec:ion}

Shortward of 1100\,\AA, the continuum of \ton\ undergoes a sharp
fall off (see Fig.~2 in BK07), which BK07 model as dust absorption
by nanodiamond grains. This resulted in a deep and broad absorption
trough that fits the observed continuum reasonably well.  In our
photoionization calculations presented below in
Sect.\,\ref{sec:pho}, we experiment with two ionizing \seds. The
first is the intrinsic `unabsorbed' \sed, which is assumed to be a
powerlaw of index $+0.1$ followed by a roll-over centered on
640\,\AA\ that extends up to the X-ray domain. Beyond 2\,\kev,
\sed\,\II\ behaves as a powerlaw of index $-1.0$, yielding an \aox\
of $-$1.45. This \sed\ is shown in Fig.\,\ref{fig:sed} and, as in
the work of BK07, it is labeled Model\,\II. The second \sed\  used
in photoionization calculations is the dust absorbed version of the
same \sed, which fits the {\it observed} UV continuum of \ton\
between 400 and 1550\,\AA\ (labeled Model\,\IV\ in
Fig.\,\ref{fig:sed}). Shortward of 200\,\AA\ and longward of
2000\,\AA, the two distributions are the same. This is because
nanodiamond dust absorbs radiation over a relatively narrow domain
as compared to other grain compositions.

\begin{figure}
\resizebox{\hsize}{!}{\includegraphics{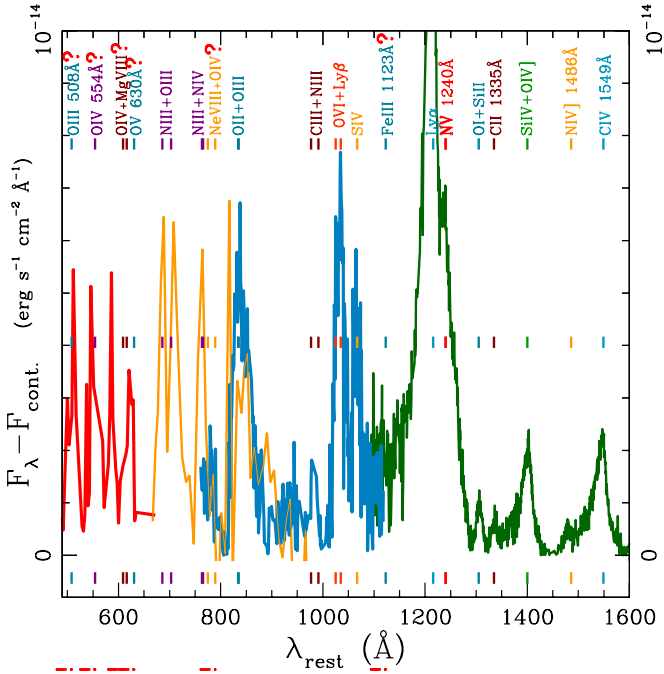}}
\caption{Residuals of the spectral energy distribution of \ton\
after subtracting our absorbed continuum Model\,\IV\ from BK07. The
different spectral segments have been color coded as follows, SWP:
red, LWP orange, HST-FOS: blue, and Palomar: dark green. Color-coded
fiducial marks indicate the position of observed or expected
(labeled with symbol '?') emission lines. Measurements of line
intensities and upper limits are given in Table\,\ref{tab:rat}.
\label{fig:res}}
\end{figure}

In Fig.\,\ref{fig:res}, we present the continuum subtracted spectrum
of \ton, that is, the residual between the observed \ton\ \sed\ and
our continuum fit represented by Model\,\IV. The different spectral
segments have been color-coded as follows, SWP: red, LWP orange,
HST-FOS: blue, and Palomar: dark green.

\subsection{Extraction of line fluxes and upper
limits}\label{sec:flu}

The procedure to measure the flux of the lines was the following: we
first fit a Gaussian to each observed line in the spectra. For
several lines, a narrow component was required, so we added a second
(narrow) Gaussian. In addition, the lines by \CIVw, \SiIVw, and
\Lya\ show a clear asymmetry in the line profile, with a blue
shoulder (see Fig.\,\ref{fig:res}). For these lines, we further
included a third, broader, Gaussian. The FWHM of the broad component
\comb{spans from $\sim3600$ to $5300$\,\kms.} It is interesting to
note that the \OII+\OIII\ complex at around 835\,\AA\ has a
significant and strong red shoulder extending up to $\sim 850$\,\AA,
which is observed in both the IUE-LWP and HST-FOS spectra (see
Fig.\,\ref{fig:res}). We could not find any positive identification
of this shoulder with any line from a different ion/transition, and
thus we considered this feature as part of the \OII+\OIII\ emission.


The measured line fluxes extracted from Fig.\,\ref{fig:res} as well
as upper limits of other permitted lines are listed with respect to
$\Lya = 100$ in Col.\,5 of Table\,\ref{tab:rat}. Note that we give
the total flux under the profile, that is, the integrated flux from
all the Gaussian components required to fit each emission line.
\comb{A consistency check was carried out, which showed that the
line fluxes measured over the original spectra or the continuum
subtracted spectra were indistinguishable from each other.}

\begin{figure}
\resizebox{\hsize}{!}{\includegraphics{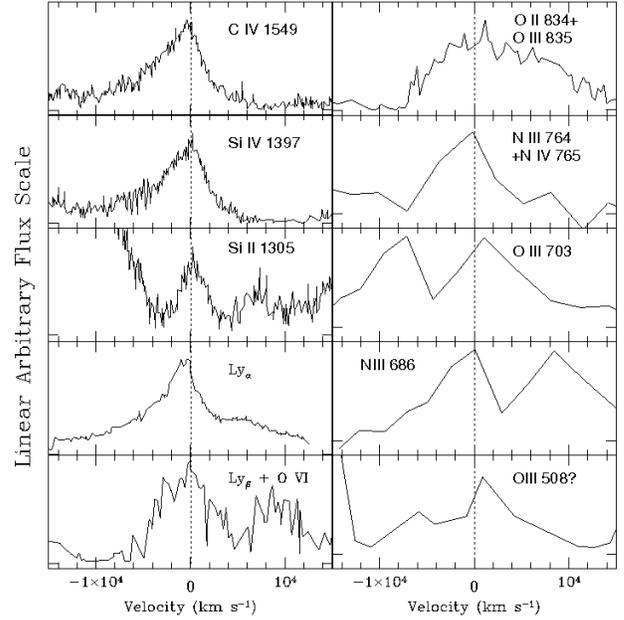}}
\caption{Emission lines extracted from the \ton\ spectrum plotted in
velocity space. The flux scale is arbitrary for each inset. Left
panels: near-UV permitted lines, right panel: far-UV permitted
lines. Overall, all the lines are consistent with the rest frame
system of \ton. Differences in the position of the lines on the
right panel may be due to absorption by intergalactic gas. The
narrow line of \OIII\ at 508\,\AA\ (bottom right panel) is severely
affected by intergalactic absorption, and better data would be
required to confirm its presence. The same applies to the other
lines shown as upper limits in Table\,\ref{tab:rat}.}
\label{fig:wid}
\end{figure}

\comb{In Col.\,5 of Table\,\ref{tab:rat}, we show our error
estimates, which we evaluated  at a 1$\sigma$ significance level. We
assumed a S/N of 25 for most lines, except for \NIII+\NIV\ and
\NIII+\OIII\ where we assume a S/N of $\simeq 10$. The line upper
limits in Table\,\ref{tab:rat} correspond to a significance of
2$\sigma$. As for the continuum, we estimate the errors to be
$\simeq 10$\%.}


Of all the emission features that we measure in the far-UV, three
line systems stand out by their strengths with respect to the
composite spectra: these are the \OII+\OIII\ lines at 835\,\AA,  the
\NIII+\OIII\ lines at 686--703\,\AA\ and the \NIII+\NIV\ lines at
765\,\AA.

Many weaker features in the IUE spectrum appear to lie where other
permitted lines of comparable excitation might be expected, such as
\OIII\ \w508\,\AA, \OIV\ \w554\,\AA, \OV\ \w630\,\AA\ and \OIV\
\w609\,\AA.  A few of these have been reported before in other
quasars (Reimers et\,al. 1998; Laor et\,al. 1995) or in composite
AGN spectra (Zheng et\,al. 1997; Telfer et\,al. 2002; Scott et\,al.
2004). \comb{However, these line systems appear as too narrow in the
IUE spectra with respect to typical BELR line profiles (see profile
comparison of Fig.\,\ref{fig:wid}).  They lack a broad component at
their base. Given the limited S/N of the IUE spectrum at the far-UV
end, we consider probable that these lines are spurious features
instead.} For this reason, we will consider these emission-like
features as upper limits rather than real detections. The symbol `?'
denotes these unconfirmed lines in our various figures.

We find little evidence of the high excitation \NeVIII\ line at
775\,\AA\ reported by Telfer et\,al. (2002) and Scott et\,al. (2004)
in their respective composite spectrum, and we favor the
identification of \OIV\ \w789\,\AA\ instead. Because the line
spectrum of \ton\ is of unusually low excitation as shown below in
Sect.\,\ref{sec:com},  we do not believe that the high excitation
lines of \MgX\ and \NeVIII\ (listed in Table\,\ref{tab:rat}) are
present at a detectable level.

\begin{figure}
\resizebox{\hsize}{!}{\includegraphics{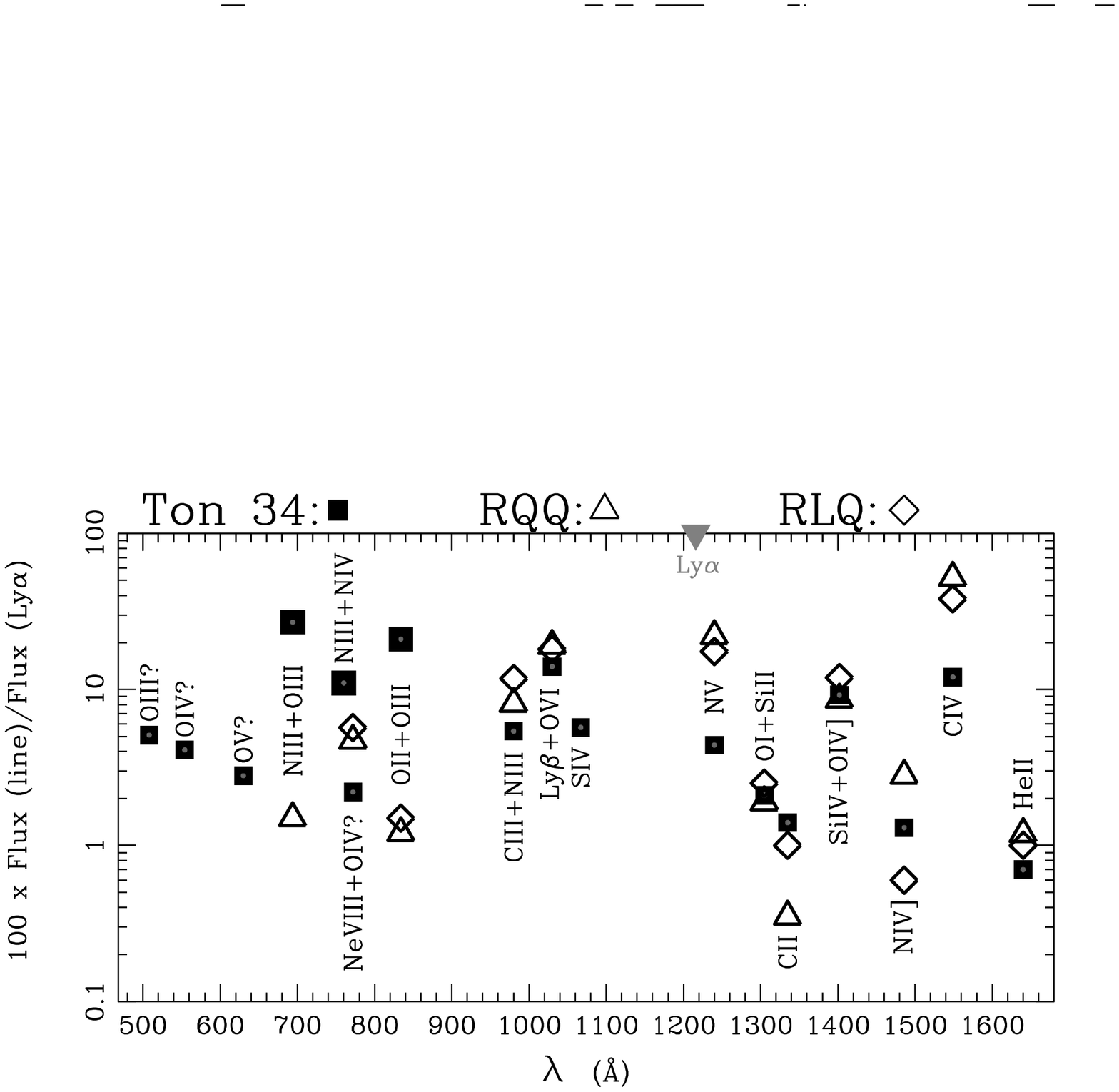}}
\caption{Line flux ratios renormalized to $\Lya = 100$ of different
species as a function of line wavelength (\AA). These were extracted
from the the spectrum of \ton\ (filled squares) and from the
radio-loud (crosses) and radio-quiet (open lozenge) composite
spectra of Telfer et\,al. (2002). Only ratios that have a
counterpart in \ton\ are shown. Larger squares correspond to ratios
for which the difference between \ton\ and the composites exceeds a
factor 10 (also shown in bold face in Col.\,5 of
Table\,\ref{tab:rat}). The symbol '?' denotes upper limits of
unconfirmed lines in \ton. The gray filled triangle indicates the
position of \Lya.} \label{fig:lin}
\end{figure}

\subsection{Originality and limitations of the data}\label{sec:lim}

As can be gathered from Fig.\,\ref{fig:res}, the strongest emission
features in the far-UV coincide with the position of lines observed
or expected in quasar spectra (Sect.\,\ref{sec:com}). However, the
limited quality of the data and the possible coincidence of
absorbers at inconvenient spectral positions prevent us from
deriving incontrovertible conclusions. In the case of the narrower
features (\OIII\ \w508\,\AA, \OIV\ \w554\,\AA, \OV\ \w630\,\AA\ and
\OIV\ \w609\,\AA), better quality data is required to confirm or
discard their presence, as discussed in Sect.\,\ref{sec:flu}.
Clearly, new observations are needed in all wave bands down to the
X-rays. In what follows, we will take the data at face value and
present photoionization and shock models that attempt to reproduce
the far-UV lines.



\begin{table*}
 \centering
\caption{Comparison of \ton\ with composite \seds\ and with models}
\label{tab:rat} \tabular{llrlllrlllrl} \hline\hline
\multicolumn{2}{c}{Lines} &  {} & \multicolumn{3}{c}{Observations} &
{} & \multicolumn{3}{c}{Photoionization$^{a}$} &  {} &
\multicolumn{1}{c}{Shocks$^{a,b}$ } \\

\cline{1-2}\cline{4-6} \cline{8-10} \cline{12-12}\\

{Species} &  {$\lambda$ (\AA)}   &   {} & {RLQ} &  {RQQ} &  {{\bf
Ton 34}} &  {} &  {KO97$^{c}$} &  {\sed\,\II$^{d}$ } &
{\sed\,\IV$^{d}$ } &  {} & {100$^{d}$\kms} \\

\hline
\\
 {(1)} &  {(2)}   &   {} &  {(3)}
&  {(4)} &  {(5)}    &   {} &  {(6)}   &  {(7)} &  {(8)} &
 {} &  {(9)} \\
 \hline
 \\

\OIII &  508 & \phn & $< 1$ & $< 1$ & $\le  5.1$ & &  ? & $10^{-2.4}$& 0.02 \phn & & 8.7 \\
\OIV &  554 & \phn & $< 1$ & $< 1$ & $\le 4.1$ & &  ? & 0.04 & 0.13 \phn & &  5.7  \\
\NeVIII &  575 & \phn &$<1$ & 2.1$^{e}$ & -- & &  ? & ? & ? \phn & &  ?  \\
\HeI & 601 & \phn &  --  &  --   &  $\le 0.5$   & & ?& 2.3 & 0.9\phn & & 2.9\\
\OIV+\MgX & 609, 617& \phn & $< 1$ & $< 1$ & $< 4.7$ & & 0.04+1.07 &0.03+ ?&0.1+ ? \phn & & 2.0+ ? \\
\OV  &  630 & \phn & $<1$ & $<1$ & $\le 2.8$ & &  0.6 & 0.13& 0.25 \phn & & 0.06 \\
\NIII+\OIII &  686, 703 & \phn & 1.5$^{e}$ & $<1$ & {\bf 27} $\pm 4.9$ & &  ?  & 0.05+0.35&  0.05+0.78\phn & & 6.8+22 \\
\NIII+\NIV & 764, 765 & \phn & $<1$ & $<1$ & {\bf 11} $\pm 1.4$ & &  0.02+0.08&  0.04+0.14&  0.04+0.04\phn &  &2.2+3.2\\
\NeVIII+\OIV &  775, 789 & \phn & 4.7 & 5.7 & $\le 2.2$ & & 2.4+1.1
& ? +0.67& ? +1.6 \phn &  & ? +5.4\\
\OII+\OIII &  834, 835 & \phn & 1.2 & 1.5 & {\bf 21} $\pm 0.32$ & &  ? +0.5& 0.04+1.4& 0.04+2.4 \phn &  & 48+23\\
\CII+\NII & 906, 912 & \phn & --   &  --   &  $\le 2.7$   & & ? & $<10^{-2}$ & $<10^{-2}$\phn & & 22+4.1\\
\CIII+\NIII &  977, 991 & \phn & 8.1 & 11.7 & 5.4 $\pm 0.64$ & & 2.9+0.7 & 4.9+0.8& 3.9+0.5 \phn & &13+7.0 \\
\Lyb+\OVI & 1025, 1035 & \phn & 19.1 & 18.1 & 14 $\pm 1.75$ & &1.1+20  & 0.37+1.5& 0.36+21.5 \phn & & 2.5+$10^{-5}$\\
\CII & 1037 & \phn &  --  &   --  & blended$^{f}$    & & ?& 0.02 & 0.02\phn & & 4.1\\
\NII+\HeII &  1084 & \phn & 5.6 & 5.5 & -- & & 0.07+0.6 & 0.01+ ?&
0.01+ ?\phn &  &4.3+ ?\\
\SIV &  1067 & \phn & $<1$ & $<1$ & 5.7 $\pm 0.23$& & ?  & 0.97&0.95  \phn & & 1.6  \\
\FeIII &  1123 & \phn & 0.28 & 2.2 & -- & & 0.01 & 0.01& $<10^{-2}$ \phn & &  0.07 \\
\CIII &  1176 & \phn & 0.44 & 0.43 & -- & & 0.4 & 4.8 & 4.9\phn &  & 1.7 \\
\SIII+\SiIII &  1194, 1207 & \phn & 1.5 & 0.47 & blended$^{f}$ & & 0.04+1.0 & 0.02+0.11 & 0.02+0.06 \phn &  & 0.5+5.0\\
\Lya &  1216 & \phn & 100 & 100 & 100$^{g}$ $\pm 10.2$ & & 100  & 100&  100 \phn &  & 100 \\
\OV & 1218 & \phn &  -- &  -- &  blended$^{f}$ & & 5.3 &  2.7 & 12.8\phn & & $10^{-4}$\\
\NV &  1240 & \phn & 22.0 & 17.5 & 4.4 $\pm 0.2$ & & 3.0 & 1.5 &  6.7\phn &  &0.03\\
\SiII &  1262 & \phn & 0.27 & 0.41 & -- & & 0.08 & 0.03 & 0.05 \phn & &  1.3\\
\OI+\SiII &  1302, 1305 & \phn & 1.9 & 2.5 & 2.1 $\pm 0.24$ & & 0.07+0.03 & $10^{-5.6}$+0.01 &  $10^{-5.0}$+0.02\phn & &  $10^{-6.9}$+0.15\\
\CII &  1335 & \phn & 0.35 & 1.0 & 1.4 $\pm 0.13$ & & 0.7  & 0.63&  0.7\phn & &  44\\
\SiIV+\OIV]  & 1397, 1402 & \phn & 8.6 & 11.9 & 9.2 $\pm 0.8$ & &3.5+2.4  & 2.9+1.8&  1.1+0.8\phn &  &3.9+0.9\\
\NIV] &  1486 & \phn & 2.8 & 0.6 & 1.3 $\pm 0.2$ & & 2.6 & 3.5&  4.5\phn &  &0.3 \\
\CIV &  1549 & \phn & 52 & 38 & 12 $\pm 1.2$ & &59  & 44 & 61.4 \phn &  & 15 \\
\HeII &  1640 & \phn & 1.2 & 1 & 0.7 $\pm ?$ & &3.0  & 4.0&  9.5\phn &  & 2.1 \\
\OIII] &  1664 & \phn & 2.3 & 0.7 & -- & &7.8  & 8.1& 6.3 \phn &  & 0.7 \\
\hline\\

\endtabular
\begin{list}{}
\item{$^{a}$} {Some observational entries in Cols.\,3--5 corresponds to the sum of two
different lines. For the corresponding models in Cols.\,6--9, we
list separately each line intensity using a $+$ symbol as
separator.}
\item{$^{b}$} {Redward of 1700\,\AA\ (down to the infrared), the
100\,\kms\ shock does not generate any strong lines. For
completeness, the only other lines of significant brightness are
\SiIII\ 1896\,\AA, \CaII\ 3969\,\AA\ and \CaII\ 3934\,\AA, whose
intensities reach 3.6\%, 1.5\% and 3\% of the intensity of \Lya,
respectively. As for the (optical and UV) \FeII\ multiplet line
systems, we can't say since they are not considered by \map.
Shortward of 400\,\AA, we expect the \HeII\ lines to
be strong, with \HeII\ 304\,\AA\ reaching 80\% of \Lya.  }\\

\item{$^{c}$} {A crude model that approximates the optimally locally
emitting BELR model described by Baldwin et\,al. (1995). Each line's
peak emissivity was extracted from the grid {\sc agn4} of
photoionization calculations published by Korista et\,al. (1997)}\\

\item{$^{d}$} {These three models were computed with \map\ assuming an
initial density \nhz\ of $4\times 10^9$\,\cmc\ and solar
metallicities. The ionization parameter is 0.04 for the two
photoionized models and zero for the shock model. At these
densities, the \Lya\ luminosities per unit area of photoionized or
shocked gas are $3.2\times 10^{7}$, $4.4 \times 10^{7}$ and
$5.5\times10^{5}$\,\llu\ for models shown in Cols.\,7, 8 and 9,
respectively. These would scale approximately in proportion to \nhz.}\\

\item{$^{e}$} {Measurement by one of us (YK) using the composite spectra
lent by R. Telfer.}\\

\item{$^{f}$} {The strong neighboring lines of \Lya\ or \OVI\
makes the determination of a meaningful upper limit impossible. }\\

\item{$^{g}$} {The \Lya\ flux in \ton\ is measured to be $6.9 \times
10^{-13}\,$\llu\ corresponding to an equivalent width of 57\,\AA. }\\
\end{list}
\end{table*}

\section{Modelling of the line spectrum}\label{sec:int}

\subsection{Line ratio comparison with  composite quasar spectra}\label{sec:com}

We now quantify to which degree the emission lines differ in \ton\
from the `average' quasar. To achieve this, we list the line ratios
characterizing the radio-loud (Col.\,3) and radio-quiet (Col.\,4)
composite spectra of Telfer et\,al. (2002) in Table\,\ref{tab:rat}.
Comparison between \ton\ and these two sets of ratios require some
caution, since significant line ratio variations exist among
quasars.  For instance, Telfer et\,al. (2002) reported that the RMS
deviation of line fluxes between the different quasars amounted to
as much as 50--70\% for the strong lines of \CIVw, \OVIw\ and \Lya.
Hence, intrinsic differences of less than a factor two between the
composites and \ton\ should not be considered significant.

To facilitate the comparison of \ton\ with the two composites, we
plot their line ratios in Fig.\,\ref{fig:lin}. Inspection of the
Table\,\ref{tab:rat} or Fig.\,\ref{fig:lin} reveals that the
commonly strong BELR lines of \CIV, \NV\ and \OVI\ are all present
in \ton. Hence the apparent \comb{sharp turndown} of the ionizing UV
in the range 650--912\,\AA\ is not affecting radically the high
excitation emission lines. In particular, the \OVIw\ line is quite
strong, although not as much as in the two composites. The \CIV\ is
substantially weaker, by more than a factor of six in \ton\ with
respect to the radio-quiet composite.  Also, the line system
\CIII+\NIII\ near 980\,\AA\ is noticeably weaker, although the flux
in this line is difficult to measure accurately due to the
uncertainties introduced by the sharp continuum bent and the many
\Lya\ forest lines.

In the far-UV, we note that the intensity of the \OII+\OIII\ and
\NIII+\OIII\ systems in \ton\ are a factor of $\sim 14$ and 18
brighter, respectively, than in the RLQ composite. There is also
evidence of significant emission of \NIII\ and/or \NIV\ at 764 and
765\,\AA, which are \emph{not} detected in the composite spectra
either.





\subsection{Photoionization vs. shock excitation}\label{sec:exc}

The line spectrum of \ton\ show peculiarities that deserve further
analysis. In particular, \OII+\OIII\ (835\,\AA),  \NIII+\OIII\ lines
(686--703\,\AA) and  \NIII+\NIV\ (765\,\AA), which are  measured
with unusual strengths with respect to \Lya. Are these emission
features necessarily genuine lines? One possibility is that
extinction resonances, unaccounted for in the extinction curve used
to model the deep continuum trough (BK07), may induce features that
looked like broad emission lines. Another possibility is that \Lya\
absorbers at intervening redshifts might generate spurious emission
features by bracketing narrow continuum regions. Although we cannot
rule out either possibility with the current data, both appear
unlikely to us, on the ground that the strongest emission features
coincide quite well with the position of plausible atomic
transitions (see Fig.\,\ref{fig:res}). The two strongest line
systems of \OII+\OIII\ (835\,\AA) and \NIII+\OIII\ (686--703\,\AA)
have previously been reported in the RLQ composite, although at a
much reduced flux level. We will thus pursue our analysis under the
assumption that the observed features are real and consist of low to
intermediate excitation \emph{permitted lines}.

\subsubsection{Photoionization calculations}\label{sec:pho}

Can photoionization account for the strength of the far-UV permitted
lines? We first establish a comparison with published BELR models,
and then evaluate the impact of a strongly absorbed ionizing
continuum.

Baldwin et\,al. (1995) showed that by integrating line fluxes over a
wide range in gas density \nh\ and impinging ionizing flux
$\varphi_{\rm H}$, one obtains a much improved fit to quasar line
spectra. Such models were dubbed ``locally optimally emitting
clouds'' (LOC). Baldwin et\,al. (1995) also showed that by
preferentially selecting the optimal slab density and impinging flux
for each individual line, one can derive a line spectrum comparable
(within a factor two) to that of a true LOC model.  To derive an
approximate LOC model, we proceeded as follows. From the grid of
photoionization models published by Korista et\,al. (1997; hereafter
KO97), we extracted the highest  equivalent width found within the
plane $\varphi_{\rm H}$ vs \nh, for each line of interest.  The
particular grid that we selected was labeled {\sc agn4}\footnote{The
selected grid {\sc agn4} comprises over a 100 lines and is available
at http://www.pa.uky.edu /$\sim$korista/grids/grids.html}. It
assumes solar abundances and a   \sed\ that  was defined by KO97,
which peaks at 22\,eV. It is the closest to our \sed\,\II\ with a
18.5\,eV turnover (Fig.\,\ref{fig:sed}; see also Haro-Corzo et\,al.
2007).

The line ratios from this approximated LOC model are shown in
Col.\,6 of Table\,\ref{tab:rat}.  Unfortunately, the \NIII+\OIII\
line system (\w\w686--703\,\AA) was not part of the {\sc agn4} grid
nor the \OII\ \w834\,\AA\ line. On the other hand, the \NIII+\NIV\
system at 765\,\AA\ and the \OIII\ line at 835\,\AA\ were. The
\NIII+\NIV\ system is a factor of a few weaker than observed while
the \w835\,\AA\ \OIII\ line is predicted an order of magnitude
weaker than the observed  \OII+\OIII\  system. As we consider
unlikely that the \w834\,\AA\ \OII\ line (absent from the {\sc agn4}
grid) is stronger than \OIII, we conclude that photoionization would
have great difficulties in fitting this system. Hence, even locally
optimally emitting clouds would not be able to account for the
intensities of at least some of the far-UV lines observed in \ton.

Could the peculiar shape of the \ton\ \sed\ be responsible for the
unusual strengths of some far-UV lines? Out of curiosity, we
calculated with the multipurpose code \map\ (Ferruit et\,al. 1997;
Binette et\,al. 1989) photoionization models using \sed\,\II\ to
compare with the absorbed \sed\,\IV, characterized by the deep
trough. We assumed solar metallicities (Anders \& Grevesse 1989) and
a gas density of $4\times 10^{9}\,$\cmc. The ionization
parameter\footnote{We use the customary definition of the ionization
parameter $\up\ = \phih/c \nnh$, which is the ratio of the density
of ionizing photons impinging on the slab $\phih/c$ to the H density
at the face of the slab \nhz.
} was varied until a maximum in the \OIII]/\Hb\
($\lambda$1663/$\lambda$4861) ratio was found, which occurred at
$\up = 0.04$. The models were truncated at a depth where H is 10\%
ionized. These calculations with $\up = 0.04$ using either \sed\
\II\ or \IV\ (both plotted  in Fig.\,\ref{fig:sed}) are reported in
Cols.\,7 and 8 of Table\,\ref{tab:rat}, respectively. Because there
are fewer soft ionizing photons in \sed\,\IV, we find that the mean
energy of  the photoelectrons is twice as high as the one given by
\sed\,\II. Hence, this must result in a hotter plasma and therefore
in stronger collisionally excited lines. A comparison of the
calculated ratios between the two models and with \ton\ (Col.\,5)
reveals that, although many metal lines in Col.\,8 (\sed\,\IV) are
often stronger by a factor of a few with respect to Col.\,7
(\sed\,\II), the deep UV trough does not result in a sufficient
increase in the strengths of the \OIII+\NIII\ lines at 683,
703\,\AA\ nor of the \OII+\OIII\ lines at 835\,\AA. In conclusion,
photoionization  predicts far-UV line intensities much too weak  in
comparison with our measurements. \comb{Furthermore, making drastic
changes in the shape of the ionizing continuum does not alter this
conclusion.}

\subsubsection{Cooling shock calculations}\label{sec:sho}

In view of the difficulties of producing strong permitted lines of
\OII, \OIII\ and \NIII\ in the case of pure photoionization, we are
lead to consider whether
collisional ionization might not be more appropriate.


To investigate this possibility, we calculated with \map\ a sequence
of steady-state plane-parallel shock models with a preshock density
of $4\times 10^{9}\,$\cmc, assuming again solar metallicities. The
postshock temperatures of the different models covered the range
$1.0 \times 10^5$ -- $8\times 10^{5}$\,K, corresponding to shock
velocities of 75 to 235\,\kms. The pre-ionization state of the
shocked gas was determined self-consistently by an iterative scheme,
using the ionizing radiation produced within the cooling shock that
propagates upstream (Dopita, Binette \& Tuohy 1984).  The time
evolution of the electron and ion temperatures were followed
separately until they equalized, making use of the equilibration
timescale as defined by Spitzer (1962).  Most of the far-UV
resonance lines are emitted downstream in layers of densities in the
range $10^{10.6}$--$10^{11.3}$\,\cmc, well below the densities of
$10^{16}$ where collisional de-excitation would become a concern for
many resonance lines. The elapsed time for the shocked gas to cool
to temperatures of 8500\,K is about 10 seconds. The adiabatic
cooling and recombination of the plasma was followed in time until
the ionized fraction reached $\le 2$\%. Because the integrated
columns of the different ions are modest in shocks, line opacities
turn out negligible with respect to those of photoionized slabs. For
instance, the line center opacity of \CIIIw\ and \CIVw\ are 20 and
1, respectively, for a 100\,\kms\ shock, compared to $10^{5.3}$ and
$10^{4.9}$ for the photoionization model of Col.\,8.

\begin{figure}
\resizebox{\hsize}{!}{\includegraphics{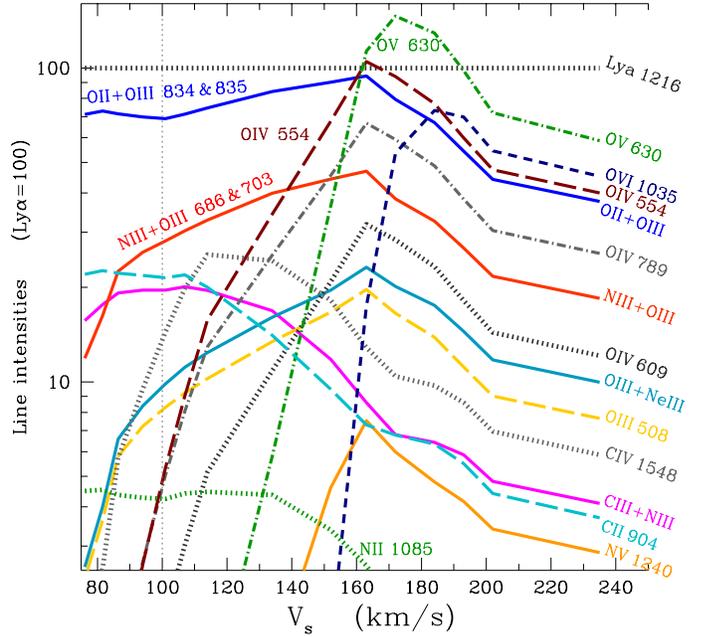}}
\caption{Line intensities from high density cooling shocks
renormalized to $\Lya = 100$ as a function of shock velocity. Solar
metallicities have been assumed. A vertical dashed line denotes the
velocity of the shock model reproduced in Table\,\ref{tab:rat}.
\label{fig:sho}}
\end{figure}

The intensities of representative far-UV lines are shown in
Fig.\,\ref{fig:sho} as a function of shock velocity.  The
calculations show that shocks with gas densities appropriate to the
BELR are very efficient in producing strong lines of \OII+\OIII\
(\w835\,\AA) and of \NIII+\OIII\ (\w\w686--703\,\AA) that reach 71\%
and 29\% of \Lya, respectively. We also computed the intensities of
many other far-UV lines that might be observable in future
observations. Some high excitation lines such as \OIV\ \w554\,\AA,
\OIV\ \w789\,\AA\ and \OV\ \w630\,\AA\ become intense for shock
velocities exceeding 120\,\kms. By comparing in Table\,\ref{tab:rat}
the observed upper limits for these lines with the computed
intensities of \OIII\ \w835\,\AA\ or \OIII\ \w703\,\AA, we find that
velocities of order 90--130\,\kms\ produce line intensities
compatible with the estimated line ratios\footnote{While the
measurements for \OIV\ \w554\,\AA\ and \OV\ \w630\,\AA\ formally
represent only upper limits, it remains possible that the
intensities of these lines are somewhat larger than evaluated given
the limited S/N of the IUE-SWP spectrum and the possible presence of
many intergalactic absorption lines (this would imply higher shock
velocities).}. To be definite, we adopt the velocity of 100\,\kms\
for the case model\footnote{For completeness, Table\,\ref{tab:rat}
includes \emph{all} the lines that the shock model predicts to be
stronger than 2\% of \Lya\ within the reported domain of
400--1700\,\AA.} presented in Col.\,9 of Table\,\ref{tab:rat}.

Shock models by themselves predict far-UV line intensities that are
too strong with respect to \Lya\ (compare Cols.\,9 and 5), creating
a reverse situation to that of photoionization
(Sect.\,\ref{sec:pho}). We are therefore lead to propose a mixed
model, in which we ascribe only a fraction of the luminosity of
\Lya\ to be due to shock excitation and the complementary fraction
to photoionization. In this mixed model, photoionization would be
responsible for the emission of the strong near-UV (i.e. classical)
lines, while shocks would be contributing of order a third of \Lya\
and (proportionally) all of the far-UV resonance lines shortward of
the Lyman limit.


The preshock density \nhz\ may be significantly higher than assumed
above. We find similar line ratios for preshock densities up to 100
times higher. The luminosity \emph{per unit area} of the shock model
in this case  exceeds that of the photoionization models presented
in Col.\,7 and 8 (see footnote {\it a} in Table\,\ref{tab:rat}). Our
code includes three body recombination of H, but not the process of
stimulated emission, which prevents us from going beyond a preshock
density of $10^{11.6}$\,\cmc. Beyond this limit, we expect \Lya\ to
be the first line to thermalize, which would further enhance the
strengths of the metal lines with respect to \Lya.

In summary, the far-UV lines observed in \ton\ shortward of the
Lyman limit are characterized by a much higher excitation energy
than the near-UV lines. For this reason,  collisional excitation
\comb{(through shocks)} at temperatures significantly \emph{higher}
than that typically provided by photoionization is strongly favored.
Calculations with \map\ show that such temperature regime is ensured
when shock excitation of moderate \Vs\ takes place. These shocks
would account not only for the far-UV lines, but may contribute
significantly to the \FeII\ multiplet lines that have been proposed
to result from mechanical heating by Joly et\,al. (2007 and
references therein).


\begin{acknowledgements}
This work was supported by the CONACyT grants J-50296 and J-49594,
and the UNAM PAPIIT grant IN118905.
Diethild Starkmeth helped us with proofreading.
\end{acknowledgements}



\clearpage

\end{document}